\documentclass[twocolumn,pre,superscriptaddress,floatfix,longbibliography]{revtex4-1}

\usepackage{graphics}
\usepackage{graphicx}
\usepackage{amsmath}
\usepackage{amssymb,xcolor}
\usepackage{subcaption} 

\newcommand{\textin}[1]{\mbox{\scriptsize{#1}}}

\definecolor{grisclair}{rgb}{0.6,0.6,0.6}

\newcommand{\ffrac}{\displaystyle\frac}

\newcommand{\beq}{\begin{equation}}
\newcommand{\ee}{\end{equation}}

\begin{document}

\title{Transient bubble rising in the presence of a surfactant at very low concentrations}
\author{D. Fern\'andez-Mart\'{\i}nez}
\address{Depto.\ de Ingenier\'{\i}a Mec\'anica, Energ\'etica y de los Materiales and\\ 
Instituto de Computaci\'on Cient\'{\i}fica Avanzada (ICCAEx),\\
Universidad de Extremadura, E-06006 Badajoz, Spain}
\author{M. G. Cabezas}
\address{Depto.\ de Ingenier\'{\i}a Mec\'anica, Energ\'etica y de los Materiales and\\ 
Instituto de Computaci\'on Cient\'{\i}fica Avanzada (ICCAEx),\\
Universidad de Extremadura, E-06006 Badajoz, Spain}
\author{J. M. L\'opez-Herrera}
\address{E.T.S.I., Depto.\ de Ingenier\'{\i}a Aeroespacial y Mec\'anica de Fluidos, Universidad de Sevilla, Camino de los Descubrimientos s/n 41092, Spain}
\author{M. A. Herrada}
\address{E.T.S.I., Depto.\ de Ingenier\'{\i}a Aeroespacial y Mec\'anica de Fluidos, Universidad de Sevilla, Camino de los Descubrimientos s/n 41092, Spain}
\author{J. M. Montanero}
\address{Depto.\ de Ingenier\'{\i}a Mec\'anica, Energ\'etica y de los Materiales and\\ 
Instituto de Computaci\'on Cient\'{\i}fica Avanzada (ICCAEx),\\
Universidad de Extremadura, E-06006 Badajoz, Spain}

\begin{abstract}
We study the formation of the dynamic adsorption layer when a bubble is released in a tank containing water with a tiny amount of surfactant. The influence of the sorption kinetic constants is examined by comparing the experiments with Sodium Dodecyl Sulfate (SDS) and Triton X-100. The experiments allowed us to determine the parameter conditions that lead to a stable bubble rising and to validate the simulation. A simple scaling analysis and the simulation show that the formation of the dynamic adsorption layer can be split into three phases characterized by disparate time scales. The mechanisms controlling those phases are surfactant convection, adsorption-desorption, and diffusion. The amount of surfactant adsorbed onto the interface increases monotonously throughout the three phases. The experiments and the simulation show that the rising velocity reaches a maximum at times of the order of $k_d^{-1}$ ($k_d$ is the desorption constant) when the dynamic adsorption layer is practically formed. This occurs even when only traces of surfactant are present in the liquid. The non-monotonous behavior of the maximum surfactant surface concentration is explained in terms of the reverse flow in the rear of the bubble right after the bubble release. This work contributes to the understanding of the complex interplay between hydrodynamics and surfactant transport and kinetics over bubble rising.
\end{abstract}

\maketitle

\section{Introduction}

Many technological and industrial processes involve the interaction between gases and liquids. Among them, we can mention waste-water treatment, chemical \citep{J14} and biochemical \citep{D13} reactors, nuclear engineering \citep{L90}, and metallurgical bubble column reactors. The heat and mass transfer across the interface of bubbles dominate the phenomena occurring in these processes. This interfacial transfer is considerably influenced by the size, shape, trajectory, and velocity of these bubbles. 

The processes mentioned above involve large populations of bubbles interacting among them. Single-bubble dynamics serve as a basis for analyzing complex multi-bubble systems. The rising of a bubble in still water is a paradigmatic problem that has been analyzed for several centuries and continues to capture the attention of many researchers today \citep{S56,SSSW08,TSG15,CTMM16,CMMT16,KJLS21,BFM23}.

When a bubble is released in a liquid bath containing surfactant, the surfactant molecules adsorb onto the free surface during the bubble rising. The molecules are advected by the outer current toward the rear of the bubble. Surfactant accumulates in that region, where desorption considerably increases. This process results in an even distribution of surfactant over the interface, producing a Marangoni stress that substantially alters the forces acting on the bubble. 

The bubble rising in the presence of a surfactant is a complex phenomenon in which fluid-dynamic and physicochemical processes are coupled, especially for large non-spherical bubbles. The bubble shape change implies variations of the interfacial area, which induces sorption processes counteracting the interface expansion or compression \citep{L62}. Interface expansion/compression, sorption kinetics, and convection over the bubble surface compete to establish the so-called {\em dynamic adsorption layer} \citep{DML98,DKGLKMM15,UGGLGM16,ZMRABSF23}.

The evolution of the bubble velocity in the presence of a surfactant significantly differs from that in clean water \citep{S56,SSSW08,TSG15,CTMM16,CMMT16,HE23,BFM23}. For sufficiently large surfactant concentrations, the bubble velocity reaches its maximum value after initial acceleration. Then, the bubble decelerates until a plateau value (the terminal velocity) is attained. This non-steady rising process corresponds to the major part of the bubble trajectory in many practical situations. The non-monotonous dependence of the bubble velocity on the vertical position (time) is usually referred to as the {\em local velocity profile} \citep{KM02,KZM07,KF10,DKGLKMM15}. 

The local velocity profile of a rising bubble is a fingerprint of the dynamic adsorption layer's transient behavior. It is strongly affected by the adsorption of surface-active species at the bubble surface. For this reason, the bubble path allows one to obtain information about the surfactant sorption kinetics that no other technique can provide, particularly when comprising proteins at various solvent conditions, such as pH and ionic strength. Adsorption and desorption occur at the same localization in most experimental configurations. Conversely, the surfactant is essentially adsorbed onto the front mobile part of a rising bubble, while desorption occurs mainly within the rear stagnant cap. This feature may favor the study of the adsorption kinetics of large molecules \citep{DKGLKMM15}.

There is abundant experimental information about surfactants' effect on the rising single bubble. For instance, it is well known that the height and width of the velocity profile maximum decrease as the surfactant concentration increases \citep{KZM07}. \citet{KF10} found that the instantaneous values of the bubble velocity and aspect ratio were correlated in their experiments. This suggests that surfactants can essentially affect bubble rise velocity through the bubble shape. \citet{FD96} showed that the bubble slowdown occurs when the surfactant coverage reaches approximately half the bubble surface. The steady-state velocity was independent of the surfactant concentration for sufficiently large concentrations \citep{SGF96,ZF01}.

Several studies have examined the effects of specific surfactants. For instance, adding a surfactant to $\beta$-lactoglobulin \citep{UKLDKJGGMM14} solutions or varying the pH of bovine serum albumin solutions \citep{ZTOEM10} dramatically alter the velocity profile of a rising bubble. The time for establishing an immobile rigid surface layer at the rising bubble surface becomes shorter with increasing pH \citep{UGGLGM16}. The bubble mobility is more influenced than the bubble shape when adding sodium dodecyl sulfate (SDS) to a pseudo-plastic liquid \citep{TKMB04}. \citet{TTM14} studied the bubble path instability in 1-pentanol and Triton X-100 solutions, observing that the drag force monotonically increases with the surfactant concentration, while the lift force showed a non-monotonic behavior. Cetyltrimethyl ammonium bromide (CTAB) dissolved in ultrapure water exhibits two distinct velocity stages, whereas Tween 80 solutions display a three-stage velocity profile \citep{LWZGZXLL22}.

The information in all the experimental studies is limited to the time-dependent bubble shape and velocity. Interpreting these experimental data requires theoretical models to describe the unsteady dynamic adsorption layer. 

Approximate analytical methods have been proposed to describe the unsteady dynamic adsorption layer and its influence on bubble rising \citep{ZKDM00}. The results only apply to spherical shapes and low Reynolds numbers \citep{ZKDM00}. In the quasi-steady approximation, the instantaneous velocity has been assumed to be the steady velocity corresponding to the amount of the solute accumulated at that instant \citep{DKGLKMM15}. Previous studies have used this theory to obtain the parameters of adsorption-desorption kinetics from the local velocity profile measured in the experiments. For non-spherical bubbles and large Reynolds numbers, accurate transient numerical simulations are required to produce reliable results.

Most numerical studies focus mainly on the final stationary (stable) \citep{M96,LE10,LM01,T05,TDM08,KADL23} or oscillating (unstable) \citep{PWMB18,FPSD24} regime reached by the rising bubble. The initial stage following the bubble release has received less attention. Transient axisymmetric simulations show that the bubble attains a maximum velocity before slowing down to its steady-state velocity \citep{LM00}. A speed comparable to the steady-state speed in pure water is reached even if the bubble is contaminated before releasing it \citep{LM00}. The elasticity number and the bulk Peclet number significantly affect the initial transient motion as well \citep{TDM08}. 

\citet{CMS97} considered a simplified dynamic problem in which the bubble remained spherical and its velocity was constant throughout the contamination process, focusing on the physicochemical processes neglected in many previous approaches. They described the temporal evolution of the relevant interfacial quantities to show the formation of the dynamic adsorption layer. The rising of a bubble covered by Triton X-100 was simulated by decoupling the solution of the fluid flow and mass transfer under different approximations \citep{ZMF01}. The best results were obtained assuming the stagnant cap model and surfactant transfer to the interface controlled by diffusion. \citet{MUT06} showed the influence of the adsorption and desorption constants on the local velocity profile and the temporal evolution of the drag coefficient. The height and width of the velocity profile maximum decrease as the adsorption (desorption) constant increases (decreases).

\citet{PWMB18} conducted three-dimensional direct numerical simulations of a bubble rising in a liquid containing a soluble surfactant. The numerical results agreed with their experiments \citep{PMKUMB17}. They concluded that the initial transient stage is very sensitive to the initial surface concentration. The quasi-steady state of the rise velocity is reached without adsorption and desorption being necessarily in equilibrium.

\citet{RVCMLH24} have recently explained how traces of surfactant can significantly change the steady (terminal) regime of bubble rising. The tiny surface tension variation occurs within an extremely thin, diffusive surface boundary layer. This produces a Marangoni stress confined within that layer that immobilizes the free surface and drastically increases the outer viscous stress only in the diffusive layer. The dynamic adsorption layer described by \citet{RVCMLH24} qualitatively differs from those described in other works under different conditions, in which the 
Marangoni and outer viscous stresses are spread over a considerable portion of the bubble surface. 

In this work, we will study numerically how the singular dynamic adsorption layer appearing at very low surfactant concentrations \citep{RVCMLH24} grows over time. We will analyze the temporal evolution of the relevant interfacial quantities and explain the effect of surfactant convection, kinetics, and diffusion on the growth of the dynamic adsorption layer. 

The numerical results of \citet{MUT06} for deformable bubbles are restricted to small Reynolds and Peclet numbers. Our boundary-fitted numerical method is similar to that employed in that work. We use a spectral collocation technique \citep{K89} to accumulate the grid points next to the interface, which allows us to resolve the extremely thin surfactant boundary layer on the outer side of the bubble surface. In this way, we solve the complete model \citep{MUT06} even for realistic values of the Reynolds and Peclet numbers. 

It is well known that the surfactant monolayer enhances the path instability, reducing the critical radius above which helical and zig-zagging trajectories are observed \citep{TTM14,RVCMLH24}. Non-axisymmetric instabilities are not allowed in our axisymmetric simulations. For this reason, we will conduct experiments to determine the parameter conditions leading to a straight path (axisymmetric flow) and to validate our numerical solutions for those conditions.
 
\section{Methods}
\label{sec2}

\subsection{Governing equations}


Consider a bubble of radius $R=[3V/(4\pi)]^{1/3}$ ($V$ is the bubble volume), density $\rho^{(i)}$, and viscosity $\mu^{(i)}$ rising in a liquid of density $\rho^{(o)}$ and viscosity $\mu^{(o)}$. The surface tension of the clean interface is $\sigma_c$, while the gravity acceleration is $g$. 

We dissolve a surfactant in the liquid at the concentration $c_{\infty}$. In the framework of the model considered here, the surfactant properties are the volumetric diffusion coefficient ${\cal D}_o$, the surface diffusion coefficient ${\cal D}_s$, the adsorption and desorption constants, $k_a$ and $k_d$, and the maximum packing density $\Gamma_{\infty}$.

\begin{figure}
\vspace{0.cm}
\begin{center}
\resizebox{0.335\textwidth}{!}{\includegraphics{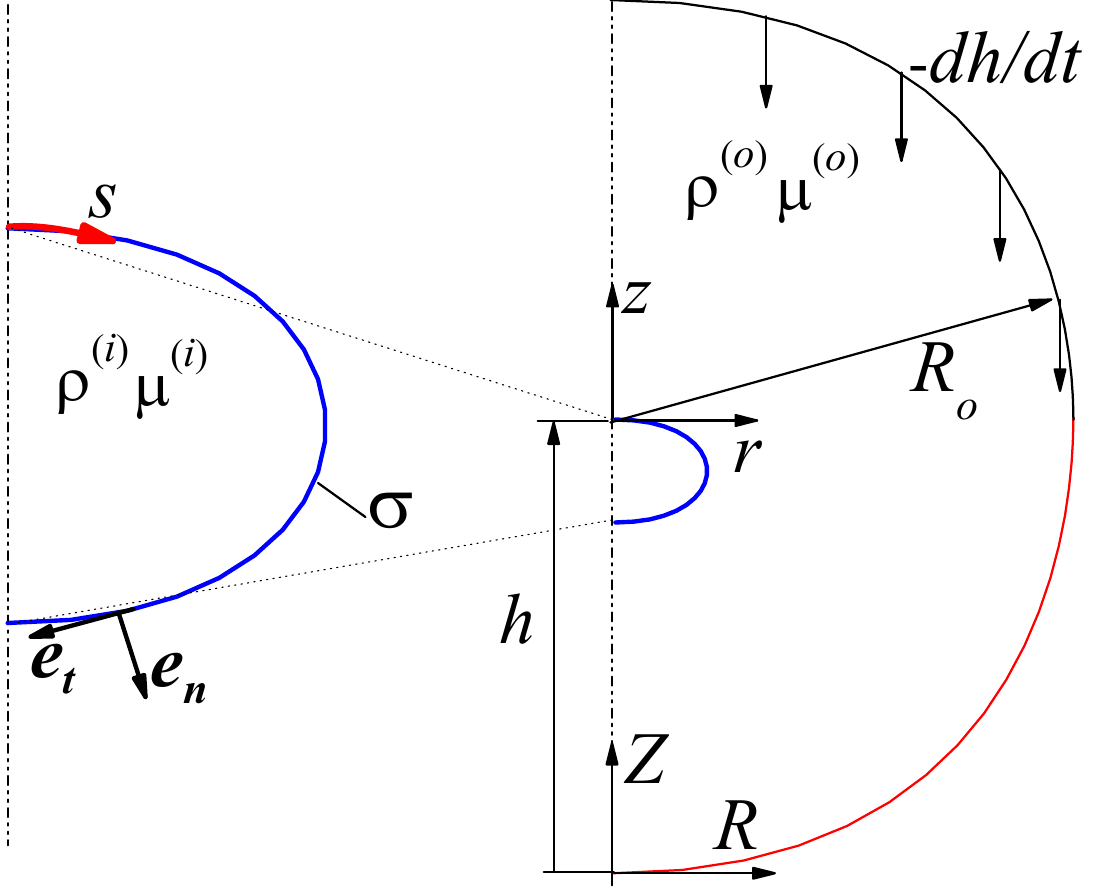}}
\end{center}
\caption{Sketch of the numerical domain. The blue and red outer boundaries correspond to the inlet and non-reflecting boundary conditions, respectively.}
\label{sketch}
\end{figure}

The hydrodynamic equations are solved in a cylindrical system of coordinates $(r,z)$ whose origin solidly moves with the bubble's upper point (Fig.\ \ref{sketch}). The continuity and momentum equations are
\begin{equation}
\label{equ1}
\boldsymbol{\nabla}\cdot\mathbf{v}^{(j)}=0, 
\end{equation}
\begin{equation}
\rho^{(j)}\frac{D\mathbf{v}^{(j)}}{Dt}=-\rho^{(j)} \left(g+\frac{d^2h}{dt^2}\right)\, \mathbf{e_z}+\boldsymbol{\nabla}\cdot\boldsymbol{\sigma}^{(j)},
\end{equation}
where $\mathbf{v}^{(j)}=u^{(j)}\mathbf{e_r}+w^{(j)}\mathbf{e_z}$ is the axisymmetric velocity field, the superscripts $j=i$ and $o$ refer to the inner and outer phases, respectively, $\mathbf{e_r}$ and $\mathbf{e_z}$ are the unit vectors along the axis $r$ and $z$, respectively, $D/Dt$ is the material derivative, $h=Z-z$ is the vertical position of the bubble's upper point (Fig.\ \ref{sketch}),
\begin{equation}
\boldsymbol{\sigma}^{(j)}=-p^{(j)} {\bf I}+\boldsymbol{\tau}^{(j)}
\end{equation}
is the stress tensor, $p^{(j)}$ is the hydrostatic pressure, ${\bf I}$ is the identity matrix, and
\begin{equation}
\boldsymbol{\tau}^{(j)}=\mu^{(j)}\left[\boldsymbol{\nabla}{\bf v}^{(j)}+\left(\boldsymbol{\nabla}{\bf v}^{(j)}\right)^T\right]
\end{equation}
is the viscous stress tensor. 

The velocity field is continuous ($\mathbf{v}^{(i)}=\mathbf{v}^{(o)}$) at the free surface. The interface is parametrized in terms of the meridional arc length $s$ ($0\leq s\leq s_f$) as $r_s=f(s)$ and $z_s=g(s)$, where $(r_s,z_s)$ is the interface location. Here, $s_f$ is the arc length corresponding to the bubble's rear point (Fig.\ \ref{sketch}). The kinematic compatibility condition reads
\begin{equation}
\frac{\partial f}{\partial t}+u^{(i)}g'-w^{(i)}f'=0,\label{kinematic}
\end{equation}
where the apostrophe $(')$ indicates the derivative with respect to $s$. 

The equilibrium of normal and tangential stresses on the interface leads to the equations
\begin{equation}
\label{bcb}
\mathbf{e_n}\cdot\left({\boldsymbol \sigma}^{(o)}-\boldsymbol{\sigma}^{(i)}\right)\cdot \mathbf{e_n}=\sigma \kappa,\quad
\mathbf{e_t}\cdot\left({\boldsymbol \sigma}^{(o)}-\boldsymbol{\sigma}^{(i)}\right)\cdot \mathbf{e_n}=\tau_{\textin{Ma}},
\end{equation}
where $\sigma$ is the local value of the interfacial tension, $\kappa=\nabla\cdot \mathbf{e_n}$ is (twice) the mean curvature, and
\begin{equation}
\mathbf{e_n}=\frac{g' {\bf e}_r-f'{\bf e}_z}{(f'^2+g'^2)^{1/2}} \quad \text{and} \quad  \mathbf{e_t}=\frac{f'{\bf e}_r+g'{\bf e}_z}{(f'^2+g'^2)^{1/2}}
\end{equation}
are the unit vectors normal and tangential to the interface, respectively. In addition, $\tau_{\textin{Ma}}=\sigma'$ is the Marangoni stress. We neglect the viscous surface stresses in Eqs.\ (\ref{bcb}) because the shear and dilatational viscosities of the surfactant monolayer are very small \citep{PRHEM20}.

We restrict ourselves to low surfactant concentrations, which implies that the surfactant is present as monomers. The monomer volumetric concentration $c^{(o)}({\bf r},t)$ in the outer phase is calculated from the conservation equation \citep{CMP09,KB19}
\begin{equation}
\label{c11}
\frac{\partial c^{(o)}}{\partial t}+\mathbf{v}^{(j)}\cdot \boldsymbol{\nabla}c^{(o)}={\cal D}_o\boldsymbol{\nabla}^2 c^{(o)}.
\end{equation}

Now, we model the transfer of monomers between the bulk and the bubble surface. The net sorption flux ${\cal J}= {\cal J}_a-{\cal J}_d$ is calculated as the difference between the adsorption ${\cal J}_a$ and desorption ${\cal J}_d$ flux. The kinetic model 
\begin{equation}
\label{km}
{\cal J}_a=k_a c^{(o)}_s\left(1-\frac{\Gamma}{\Gamma_{\infty}}\right),\quad\quad {\cal J}_d=k_d \Gamma  
\end{equation}
is adopted to calculate these fluxes. As mentioned above, $k_a$ and $k_d$ are the adsorption and desorption constants, respectively, $c^{(o)}_s$ is the bulk surfactant concentration evaluated at the interface, $\Gamma$ is the surfactant surface concentration (the surface coverage, measured in mols per unit area), and $\Gamma_{\infty}$ is the maximum packing density. 

The surfactant surface concentration $\Gamma$ verifies the advection-diffusion equation \citep{CMP09}
\begin{equation}
\label{ad}
\frac{\partial \Gamma}{\partial t}+{\boldsymbol \nabla}_s\cdot(\Gamma {\bf v}_s)+\Gamma({\boldsymbol \nabla}_s\cdot {\bf n})({\bf v}\cdot\mathbf{e_n})={\cal D}_s\nabla_s^2 \Gamma+{\cal J},
\end{equation}
where ${\boldsymbol\nabla_s}$ is the tangential intrinsic gradient along the free surface, ${\bf v}_s={\boldsymbol {\sf I}}_s{\bf v}$ is the (two-dimensional) surface velocity, ${\boldsymbol {\sf I}}_s={\boldsymbol {\sf I}}-\mathbf{e_n}\mathbf{e_n}$ is the tensor that projects any vector on that surface, ${\boldsymbol {\sf I}}$ is the identity tensor, and ${\cal D}_s$ is the surface diffusion coefficient.

The dependence of the surface tension $\sigma$ on the surface concentration $\Gamma$ is given by the Langmuir equation of state \citep{T97}
\begin{equation}
\label{lan}
\sigma=\sigma_c+\Gamma_{\infty} R_g T\ln \left(1-\frac{\Gamma}{\Gamma_{\infty}}\right),  
\end{equation}
where $\sigma_c$ is the surface tension of the clean interface, $R_g$ is the gas constant, and $T$ is the temperature. 

We assume that the liquid bath is not perturbed by the bubble in the upstream region $r=R_o$ and $z>0$ (Fig.\ \ref{sketch}). Therefore,
\begin{equation}
w^{(o)}=-\frac{dh}{dt}, \quad u^{(o)}=0, \quad p^{(o)}+\rho^{(o)} g z=\text{const.}
\label{entranceflow}
\end{equation}
in that boundary. The non-reflecting boundary conditions 
\begin{equation}
\frac{\partial u^{(o)}}{\partial z}=\frac{\partial w^{(o)}}{\partial z}=0
\label{outflow}
\end{equation}
are applied in the downstream region far from the bubble ($r=R_o$ and $z<0$) to capture the wake (Fig.\ \ref{sketch}). The surfactant concentration at $r=R_o$ is $c_{\infty}$.

We consider the regularity conditions 
\begin{equation}
u^{(j)}=\frac{\partial w^{(j)}}{\partial r}=\frac{\partial p^{(j)}}{\partial r}=\frac{\partial c^{(o)}}{\partial r}=0 
\end{equation}
at the symmetry axis $r=0$. The condition $w^{(o)}=0$ at the interface upper point allows us to calculate the bubble's vertical position $h(t)$. Finally, we specify the bubble's volume through the equation
\begin{equation}
V=\pi\int_0^{s_f}  f^2 g'\, ds.    
\end{equation}

We start the simulation from the hydrostatic solution corresponding to a spherical bubble. Consider a bubble ``instantaneously" inflated in a tank of still water with surfactant. The Damkohler number Da$=\tau_D/\tau_k$ relates the time scales $\tau_D$ and $\tau_k$ corresponding to the surfactant transfer limited by diffusion and sorption, respectively (see the Appendix) \citep{MS20}. Considering the values shown in the next section, the Damkohler number takes values of the order of $10$ and $10^2$ for SDS and Triton X-100, respectively, indicating that the surfactant transfer to the bubble is limited by diffusion in both cases. In the experiment, the bubble remained attached to the capillary for approximately 5 s. The surface concentrations of SDS and Triton X-100 at that instant were similar to the equilibrium concentration for SDS (see Fig.\ \ref{difu} in the Appendix). For this reason, we consider this value as the initial surface concentration in all the simulations. 

The above equations are numerically integrated with a variant of the boundary-fitted spectral method proposed by \citet{HM16a}. The major difficulty associated with a soluble surfactant is the existence of a very thin diffusive boundary layer next to the interface for the small diffusion coefficients of most surfactants (large Peclet numbers). We use Chebyshev spectral collocation points to accumulate the grid points next to the interface \citep{HM16a}, facilitating the resolution of this layer. 

\subsection{Experimental method}
\label{sec3}

In an experiment, nitrogen was injected through a needle to form the bubble in the center of the tank bottom. The bubble detached from the needle and rose across the tank until it reached the free surface. We used a virtual binocular stereo vision system \citep{LWZGZXLL22} to image two perpendicular views of the rising bubble. The images were processed at the pixel level \citep{C86} to determine the bubble shape and velocity. We show the values averaged over five experimental realizations. The experimental uncertainty corresponds to the standard deviation. Details of the experimental setup and the image processing analysis can be found in the Supplemental Material. This experimental method was validated by \citet{RVCMLH24} from comparison with the results for clean water obtained by \citet{D95}. 

We consider SDS and Triton X-100 in our experiments. SDS is an anionic surfactant with a molecular weight of 288.4 g/mol, while Triton X-100 is a nonionic surfactant with a molecular weight of 647 g/mol. Table \ref{prop} shows the relevant properties of these surfactants. The diffusion coefficients are commensurate with each other. SDS is slightly more active than Triton X-100, as indicated by the value of $\Gamma_\infty$. The SDS adsorption constant is also similar to that of Triton X-100. Interestingly, the desorption rate $k_d$ of Triton X-100 is two orders of magnitude smaller than that of SDS. For small surfactant concentrations, $\Gamma=k_a c/k_d$ at equilibrium [Eqs.\ (\ref{km})].  This means that one needs to dissolve much fewer moles of Triton X-100 in water to achieve the same surfactant surface concentration at equilibrium. This is reflected in the critical micelle concentration, around 35 times smaller in the Triton X-100 case.

\begin{table*}
\begin{tabular}{ccccccc}
&CMC  (mol/m$^3$)&$k_a$ (m/s)& $k_d$ (s$^{-1}$) &$\Gamma_\infty$ (mol/m$^2$)&${\cal D}_s$ (m$^2$/s)&${\cal D}_o$ (m$^2$/s)\\ \hline
SDS & 8.0 & $2.34 \times 10^{-4}$ & 6.4 & $3.9 \times 10^{-6}$ & $4.2 \times 10^{-7}$ & $8.0 \times 10^{-10}$ \\
Triton X-100 & 0.23 & $1.45 \times 10^{-4}$ & $3.3 \times 10^{-2}$ & $2.9 \times 10^{-6}$ & $2.6 \times 10^{-10}$ & $2.6 \times 10^{-10}$ \\ \hline
\end{tabular}
\caption{Properties for SDS and Triton X-100. The values for SDS were those considered by \citet{RVCMLH24}. The values for Triton X-100 were taken from Refs.\ \citep{TTM14} and \citep{LMM90}.}
\label{prop}
\end{table*}


\section{Experimental results}
\label{sec4}

\begin{figure*}[hbt]
\includegraphics[width=0.32\textwidth]{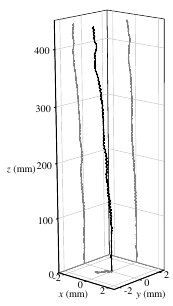}\includegraphics[width=0.32\textwidth]{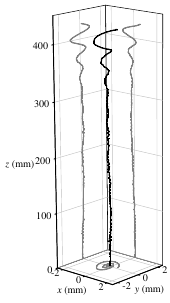}\includegraphics[width=0.32\textwidth]{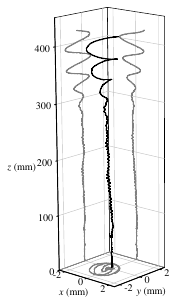}\\
\includegraphics[width=0.32\textwidth]{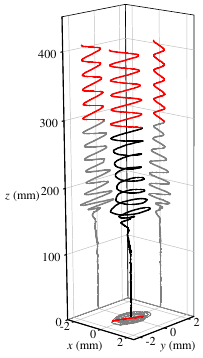}\includegraphics[width=0.32\textwidth]{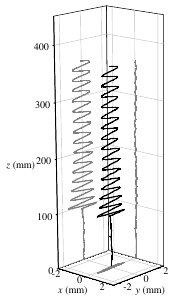}
\caption{Bubble trajectory for $R=0.76$ mm and $c_{\infty}/c_{\textin{cmc}}=10^{-4}$ (a), $5\times 10^{-4}$ (b), $10^{-3}$ (c), $3.6 \times 10^{-3}$ (d), and $10^{-2}$ (e). The graphs also show the projections of the trajectories onto the planes $(x,y)$, $(x,z)$, and $(y,z)$.} 
\label{exam}
\end{figure*}

As mentioned in the Introduction, the goal of our experimental study is twofold: (i) to determine the parameter conditions that lead to an axisymmetric, straight  (stable) bubble rising and (ii) to validate the numerical method. Figure \ref{exam} shows the three types of trajectories observed in our experiments with Triton X-100. Similar results were obtained for SDS \citep{RVCMLH24}. 

Figure \ref{exam}a corresponds to a bubble following a quasi-straight trajectory. The small amplitude oscillations can be attributed to the camera vibration while moving (we did not observe those oscillations when the camera was at rest). The path is slightly tilted (the tilt angle is smaller than $0.1^{\circ}$), probably due to a small asymmetry in the bubble formation process. For $c_{\infty}/c_{\textin{cmc}}=5\times 10^{-4}$ [case (b)], the bubble motion remains stable until $z\simeq 350$ mm. Helical instability develops at larger distances from the ejector. This experiment highlights the importance of conducting experiments with large tanks and may explain the discrepancies among the critical conditions found in previous works.  

The distance from the ejector at which the helical instability develops decreases as the surfactant concentration increases (Fig.\ \ref{exam}c). For $c_{\infty}/c_{\textin{cmc}}=3.3\times 10^{-3}$ [case (d)], the helical instability evolves toward a zig-zag motion, as shown by the projection of the bubble path onto the $xy$ plane (the red line corresponds to the zig-zag motion). The zig-zag instability arises close to the bubble ejector for the highest surfactant concentration considered in our analysis [case (e)].  

\begin{figure}[hbt]
\begin{center}
\resizebox{0.465\textwidth}{!}{\includegraphics{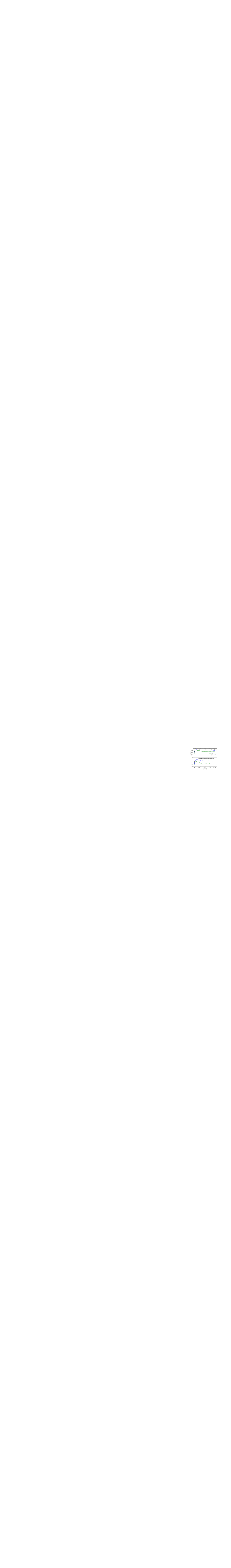}}
\end{center}
\caption{Bubble velocity $v_z$ and aspect ratio $\chi$ as a function of the vertical position $z$ of the center of gravity for $R=0.76$ mm and $c_{\infty}/c_{\textin{cmc}}=5\times 10^{-4}$. The solid and dashed lines correspond to the stable and oscillatory parts of the bubble trajectory, respectively.} 
\label{duplicity}
\end{figure}

\begin{figure}[hbt]
\begin{center}
\resizebox{0.465\textwidth}{!}{\includegraphics{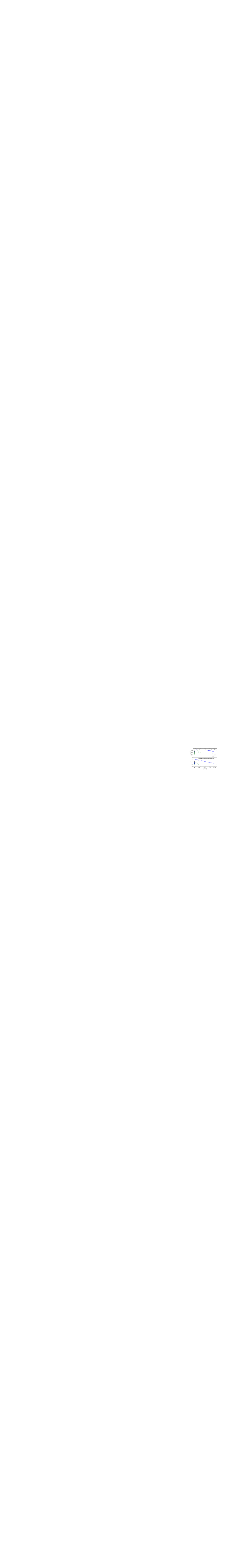}}
\end{center}
\caption{Bubble velocity $v_z$ and aspect ratio $\chi$ as a function of the vertical position $z$ of the center of gravity for $R=0.76$ mm and $c_{\infty}/c_{\textin{cmc}}=10^{-3}$. The solid and dashed lines correspond to the stable and oscillatory parts of the bubble trajectory, respectively.} 
\label{duplicity2}
\end{figure}

\begin{figure}[hbt]
\begin{center}
\resizebox{0.465\textwidth}{!}{\includegraphics{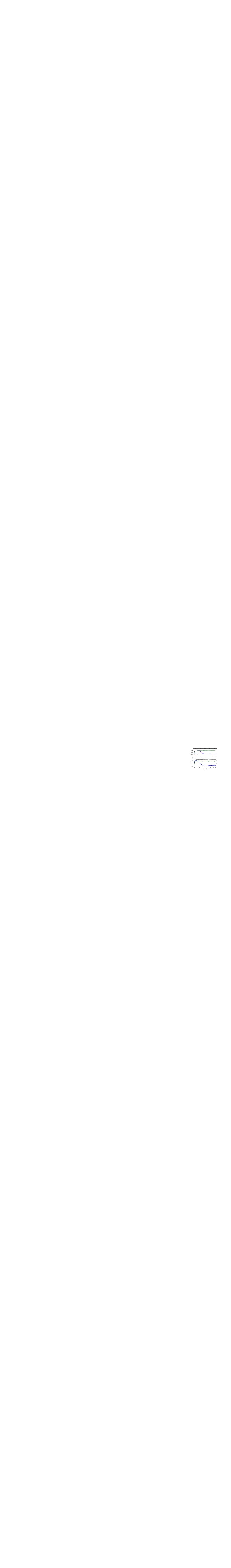}}
\end{center}
\caption{Bubble velocity $v_z$ and aspect ratio $\chi$ as a function of the vertical position $z$ of the center of gravity for $R=0.66$ mm and the concentrations $c_{\infty}=8.3\times 10^{-4}$ mol/m$^3$ of Triton X-100 and $c_{\infty}=8\times 10^{-4}$ mol/m$^3$ of SDS. The solid and dashed lines correspond to the stable and oscillatory parts of the bubble trajectory, respectively. The red circles correspond to the experimental data of \citet{ZF01} for $R=0.70$ mm and Triton X-100 at the concentration $c_{\infty}=7.5\times 10^{-4}$ mol/m$^3$.} 
\label{absolute2}
\end{figure}

Figures \ref{duplicity}--\ref{absolute2} compare the effect of SDS and Triton X-100 on the bubble velocity and aspect ratio. We monitor the bubble velocity and the aspect ratio to safely determine whether the bubble has reached a steady motion. 

The results in Figs.\ \ref{duplicity} and \ref{duplicity2} were obtained for the same value of the relative concentration $c_{\infty}/c_{\textin{cmc}}$ of SDS and Triton X-100. Equations (\ref{km}) yield $\Gamma_{\textin{eq}}/\Gamma_{\infty}=[k_a c_{\textin{cmc}}/(k_d \Gamma_{\infty})]\, c/c_{\textin{cmc}}$  at equilibrium, where $k_a c_{\textin{cmc}}/(k_d \Gamma_{\infty})=75$ and 348 for SDS and Triton X-100, respectively. This means that the same value of $c/c_{\textin{cmc}}$ leads to a much higher equilibrium surface coverage $\Gamma_{\textin{eq}}/\Gamma_{\infty}$ and density $\Gamma_{\textin{eq}}$ in the Triton X-100 case. However, the Triton X-100 equilibrium surface coverage is reached at much longer times due to the much smaller value of the desorption constant. This suggests that Triton X-100 must produce a smaller effect on a relatively short time scale, but this effect must eventually exceed that produced by SDS at sufficiently large times. The results in Figs.\ \ref{duplicity} and \ref{duplicity2} are consistent with the above prediction. The bubble velocity and aspect ratio for SDS are smaller than those for Triton X-100. Nevertheless, a crossover is expected to occur beyond the maximum height analyzed in the experiment.  

\begin{figure}[hbt]
\begin{center}
\resizebox{0.465\textwidth}{!}{\includegraphics{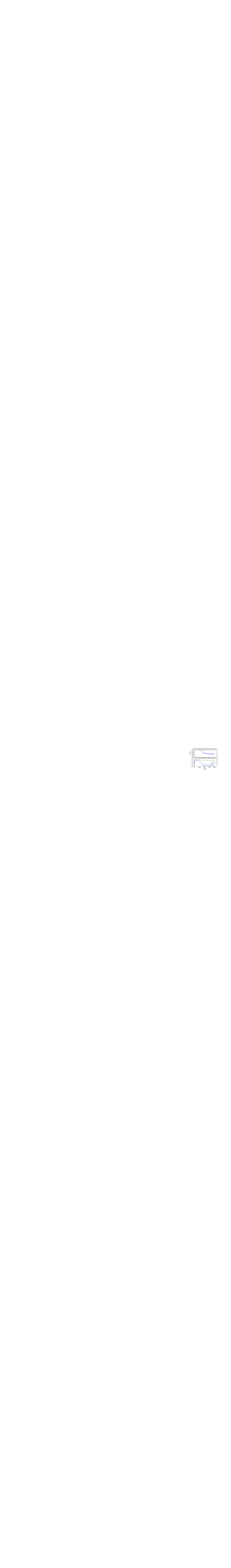}}
\end{center}
\caption{Bubble velocity $v_z$ and aspect ratio $\chi$ as a function of the vertical position $z$ of the center of gravity for $R=0.76$ mm and the concentrations $c_{\infty}=8.3\times 10^{-4}$ mol/m$^3$ of Triton X-100 and $c_{\infty}=8\times 10^{-4}$ mol/m$^3$ of SDS. The solid and dashed lines correspond to the stable and oscillatory parts of the bubble trajectory, respectively.} 
\label{absolute}
\end{figure}

The results in Figs.\ \ref{absolute2} and \ref{absolute} were obtained for practically the same value of the absolute concentration $c_{\infty}$ of SDS and Triton X-100. In this case, the adsorption flux ${\cal J}_a\simeq k_a c_{\infty}$ is expected to take similar values for the two surfactants. This explains why the bubbles covered with SDS and Triton X-100 exhibit similar velocities and aspect ratios in the first stage of the bubble rising. As explained in Sec.\ \ref{sec3}, the same volumetric concentration leads to a much higher Triton X-100 surface concentration at equilibrium. This explains why the Triton X-100 effects are larger than those caused by SDS at the same absolute concentration (Figs.\ \ref{absolute} and \ref{absolute2}).

Figures \ref{absolute2} and \ref{absolute} allow us to analyze the influence of the bubble radius. The surfactant effect increases as the bubble radius decreases. Specifically, the terminal velocity reduction caused by SDS is larger for the $R=0.66$ mm case, and Triton X-100 destabilizes the bubble path at smaller values of $z$ in that case. This result can be expected because the surfactant produces interfacial effects, and the surface-to-volume ratio increases as $R$ decreases.

All the trajectories analyzed in Figs.\ \ref{duplicity}--\ref{absolute2} become unstable except for the case SDS at $c_{\infty}=8\times 10^{-4}$ mol/m$^3$, where the bubble reaches a steady state. In the next section, we numerically analyze the temporal evolution of the dynamic adsorption layer for that case. The analysis of this axisymmetric flow is realistic because the flow does not suffer from non-axisymmetric instabilities over the entire bubble trajectory.

\section{Numerical results}
\label{sec5}

\subsection{Validation of the numerical method}
\label{sec51}

We calculated the numerical solution for $\rho^{(i)}=1.2$ kg/m$^3$, $\rho^{(o)}=997$ kg/m$^3$, $\mu^{(i)}=1.83\times 10^{-5}$ kg/(m$\cdot$s), $\mu^{(o)}=9.3\times 10^{-4}$ kg/(m$\cdot$s), and $\sigma_c=72$ mN/m. The values of $\Gamma_{\infty}$, ${\cal D}_o$, and  ${\cal D}_s$ were taken from Table \ref{prop}. The values of these surfactant properties are realistic except for the surface diffusion coefficient ${\cal D}_s$, which is significantly smaller than that considered in our simulations. We increased the value of ${\cal D}_s$ to favor the convergence of the numerical method. We verified that an increase in ${\cal D}_s$ of the order of $10^{-1}$ produces an increase in the terminal velocity of the order of $10^{-5}$. This occurs because the Peclet number is sufficiently large for the surfactant surface distribution to be dominated by the advection and adsorption-desorption kinetics \citep{TTM14}.

The reported values of the adsorption and desorption constants of most surfactants are relatively inconsistent. For this reason, these values are adjusted in the simulation to reproduce the bubble path observed experimentally. In fact, bubble rising has been proposed to determine the values of those constants. We followed different strategies to determine the adsorption and desorption constants of Triton X-100 and SDS. 

For the small surfactant concentrations considered in our analysis, $\Gamma/\Gamma_{\infty}\ll 1$, which implies that
\begin{equation}
\sigma_c-\sigma\simeq  L_d\, c_{\infty}\, R_g\, T
\end{equation} 
at equilibrium. This equation shows the critical role played by the depletion length $L_d=k_a/k_d$ (or, equivalently, the Langmuir equilibrium adsorption constant $K_L=L_d/\Gamma_{\infty}$) in the bubble dynamics. A reliable value of the depletion length, $L_d=4.4$ mm, has been determined for Triton X-100 in previous studies \citep{PF01,TTM14}. For this reason, it is reasonable to fit the numerical bubble path to the experimental one by varying $k_a$ and $k_d$ while keeping $L_d$ equal to that value \citep{LMM90,FD96}. We have done this with Triton X-100.

The surface tension measurement for SDS does not lead to a reliable value of $L_d$ \citep{FD96,PF01}. Alternatively, \citet{RVCMLH24} took the desorption constant $k_d$ from accurate experimental measurements and used $k_a$ as the fitting parameter, whose experimental value exhibits more uncertainty. The value shown in Table \ref{prop} led to the best agreement between the numerical terminal velocity and the experimental one for the case analyzed in this work. We take that value for the transient simulations conducted here. The depletion length $L_d$ obtained from this procedure is consistent with the value given by \citet{FD96}.

\begin{figure}[hbt]
\begin{center}
\resizebox{0.465\textwidth}{!}{\includegraphics{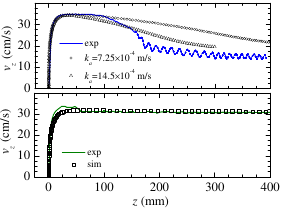}}
\end{center}
\caption{Bubble velocity $v_z$ as a function of the vertical position $z$ of the center of gravity for Triton X-100 (upper graph) and SDS (lower graph). The experiments with Triton X-100 were conducted for $R=0.76$ mm and $c_{\infty}=8.3\times 10^{-4}$ mol/m$^3$. The experiments with SDS were obtained for $R=0.66$ mm and $c_{\infty}=8\times 10^{-4}$ mol/m$^3$. The solid and dashed lines correspond to the stable and oscillatory parts of the experimental bubble trajectory, respectively. The symbols are the simulation results. The simulation results for Triton X-100 were calculated with $k_a=7.25\times 10^{-4}$ m/s and $14.5\times 10^{-4}$ m/s. In the two cases, $L_d=4.39$ mm. The results for SDS were obtained with the values of $k_a$ and $k_d$ shown in Table \ref{prop}.} 
\label{absolute3}
\end{figure}

Figure \ref{absolute3} compares the numerical and experimental local velocity profiles for Triton X-100 and SDS. The numerical results for Triton X-100 were calculated for $k_a=7.25\times 10^{-4}$ m/s and $14.5\times 10^{-4}$ m/s while keeping $L_d$ constant ($L_d=4.4$ mm), as explained above. The simulation with $k_a=14.5\times 10^{-4}$ m/s approximately captures the dependency of the bubble velocity on the vertical coordinate. Significant deviations arise in the oscillatory part of the experimental bubble trajectory, which suggests that the oscillatory instability reduces the average vertical velocity. 

The results for SDS show a good agreement between the simulation and the experiment. For $z\gtrsim 80$ mm, both the bubble velocity and the aspect ratio take approximately constant values (Fig.\ \ref{absolute2}), indicating that a quasi-steady regime has been reached. In this regime, the bubble velocity is significantly smaller than that in pure water. This difference is accurately captured by the simulation. The experimental local velocity profile exhibits the so-called ``overshooting"\ phenomenon: the velocity reaches a maximum and decreases to its terminal value. This phenomenon is less noticeable in the simulation, which suggests that the formation of the numerical adsorption layer is slightly delayed with respect to its experimental counterpart, as shown in the next section.

\subsection{Growth of the dynamic adsorption layer}
\label{sec52}

This section analyzes numerically the growth of the dynamic adsorption layer formed when a bubble is released in water containing a surfactant at a very low concentration. Specifically, we consider the same configuration as that studied by \citet{RVCMLH24} in the steady regime, i.e., a bubble with $R=0.66$ mm in water containing SDS at the concentration $c_{\infty}=8\times 10^{-4}$ mol/m$^3$ ($c_{\infty}/c_{\textin{cmc}}=10^{-4}$) (see Figs.\ \ref{absolute2} and \ref{absolute3}). 

\begin{figure}[tbh]
\begin{center}
\resizebox{0.37\textwidth}{!}{\includegraphics{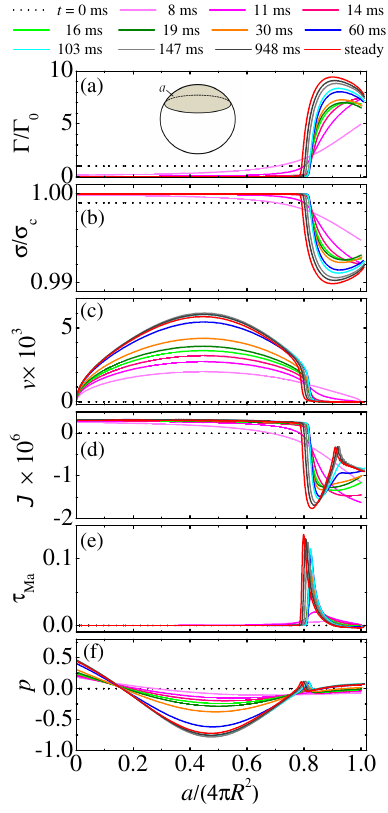}}
\end{center}
\caption{Surfactant surface concentration $\Gamma$, surface tension $\sigma$, surface velocity $v$, net sorption flux ${\cal J}={\cal J}_a-{\cal J}_d$, Marangoni stress $\tau_{\textin{Ma}}$, and hydrostatic pressure $p$ as a function of the area $a$ for $R=0.66$ mm and $c_{\infty}=8\times 10^{-4}$ mol/m$^3$ of SDS. The velocity, flux and stresses are measured in terms of $v_{\sigma\mu}$, $v_{\sigma\mu}\Gamma_{\infty}/R$ and $\rho^{(o)} v_{\sigma\mu}^2$, respectively, where $v_{\sigma\mu}=\sigma_0/\mu^{(o)}$ is the visco-capillary velocity.} 
\label{Profiles_time}
\end{figure}

Figure \ref{Profiles_time} shows the surfactant monolayer evolution during the bubble rising. Since the bubble shape changes over time, the polar angle measured from the bubble's center of gravity also changes, and its use as the independent variable is confusing. For this reason, we have selected instead the surface area $a$ measured from the north pole. The surfactant surface concentration $\Gamma$ sharply increases in the rear of the bubble. The maximum of $\Gamma(a)$ displaces toward the bubble equator until it reaches the approximately constant location $a/(4\pi R^2)\simeq 0.8$, where $\Gamma$ becomes almost ten times the initial concentration $\Gamma_0$. This large increase in $\Gamma$ occurs within a very thin diffusive layer, producing a significant surface tension gradient even though this quantity changes in less than 1\%. The surface tension gradient gives rise to a Marangoni stress $\tau_{\textin{Ma}}$ three orders of magnitude larger than the tangential viscous stress in a surfactant-free bubble \citep{RVCMLH24}. Although the Marangoni stress is confined within the surface boundary layer, it immobilizes part of the bubble’s south hemisphere and significantly reduces the terminal velocity (Fig.\ \ref{absolute2}). 

\begin{figure}
\begin{center}
\resizebox{0.425\textwidth}{!}{\includegraphics{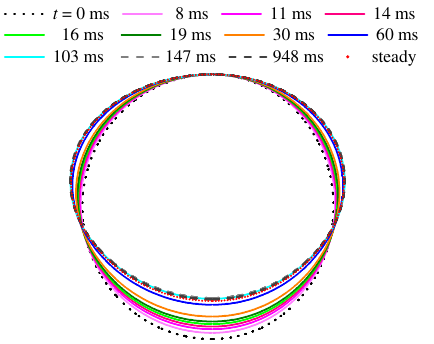}}
\end{center}
\caption{Evolution of the bubble shape for $R=0.66$ mm and $c_{\infty}=10^{-4}$ mol/m$^3$ of SDS.} 
\label{Shape_evol}
\end{figure}

The results for the net sorption flux ${\cal J}={\cal J}_a-{\cal J}_d$ show that surfactant adsorbs onto (desorbs from) the interface in front of (behind) the diffusive layer over the entire bubble motion. The hydrostatic pressure distributions over the bubble surface reflect the bubble acceleration. Pressure is built up in front of the bubble as the bubble velocity increases, while a negative gauge pressure arises around the equator, flattening the bubble (Fig. \ref{Shape_evol}).    

\begin{figure}[hbt]
\begin{center}
\resizebox{0.4\textwidth}{!}{\includegraphics{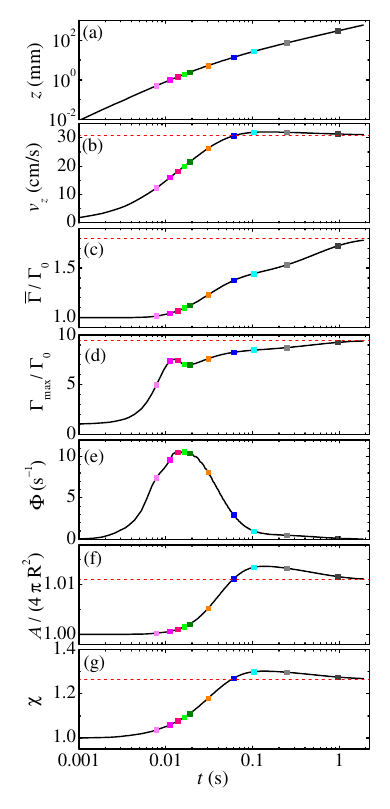}} 
\end{center}
\caption{Bubble vertical position $z$, velocity $v_z$, mean surfactant concentration $\overline{\Gamma}$, maximum surfactant concentration $\Gamma_{\textin{max}}$, normalized net sorption rate $\Phi$, bubble surface area $A$, and aspect ratio $\chi$ as a function of time $t$ for $R=0.66$ mm and $c_{\infty}=8.0\times 10^{-4}$ mol/m$^3$ of SDS. The symbols indicate the instants corresponding to the profiles in Fig.\ \ref{Profiles_time}. The dashed horizontal lines indicate the terminal values.} 
\label{evolution}
\end{figure}

Figure \ref{evolution} shows the evolution of the main quantities describing the bubble dynamics. The symbols indicate the instants corresponding to the profiles in Fig.\ \ref{Profiles_time}. In clean water, the bubble velocity monotonously increases until it asymptotically reaches its terminal value. However, the velocity in our simulation reaches a maximum at $t\sim 0.1$ s and then slightly decreases (Fig.\ \ref{evolution}b). This decrease is produced by the extra drag resulting from the growth of the dynamic surfactant layer. The wake behind the bubble is almost at rest due to the interface immobilization. This reduces the pressure in that region, which increases the drag. The sharp increase in the Marangoni stress causes a peak in the normal outer viscous stress that also contributes to the drag \cite{RVCMLH24}. These two effects are responsible for the bubble deceleration for $t>0.1$ s. 

The mean surfactant concentration $\overline{\Gamma}$ monotonously increases (Fig.\ \ref{evolution}c) due to the continuous net adsorption of surfactant. As explained below, the sharp increase in $\Gamma_{\textin{max}}$ for $t\lesssim 0.01$ s (Fig.\ \ref{evolution}d) reflects the intense surfactant surface convection right after the bubble release. To analyze this aspect of the problem, we calculate the normalized net sorption rate
\begin{equation}
\label{Phi}
\Phi=\frac{1}{4\pi R^2 \Gamma_0}\frac{d}{dt}\left(A \overline{\Gamma}\right),
\end{equation}
where $A$ is the bubble surface area. $\Phi$ is the time derivative of the surfactant mass trapped in the bubble, $A \overline{\Gamma}$, relative to its initial value. As explained below, it reaches its maximum value at $t\simeq 0.016$ s (Fig.\ \ref{evolution}e), when the bubble front (clean) surface has grown. Figure \ref{evolution}f shows that the bubble surface slightly expands ($A$ increases) during the bubble acceleration and compresses during the deceleration. This expansion/compression is accompanied by a significant change in the aspect ratio $\chi$ (Fig.\ \ref{evolution}g), as also observed in Fig.\ \ref{Shape_evol}. The instantaneous values of the bubble velocity and aspect ratio are correlated, as \citet{KF10} observed in their experiments.

Four time scales describe the formation of the dynamic adsorption layer. The fastest process is convection to the rear part of the bubble of the surfactant adsorbed before the bubble is released. The time scale of this mechanism is $t_c=R/v_{zc}\sim 10^{-2}$ s, where $v_{zc}\simeq 0.1$ m/s is the characteristic rising velocity during the bubble acceleration. This first process is followed by that influenced by adsorption-desorption kinetics. The adsorption time scale is $t_a=(k_a c_s/\Gamma)^{-1}$. The initial surfactant concentration is approximately that at equilibrium. Then, $c_s/\Gamma\simeq k_d/k_a$, and, therefore, $t_a=k_d^{-1}$. We conclude that the adsorption time scale is commensurate with the desorption characteristic time $t_d=k_d^{-1}\sim 0.1$ s, which suggests that both mechanisms compete with each other during the same phase of the dynamic adsorption layer formation. The surfactant diffusion governs the last stage of this process. The surface and volumetric diffusion times are $t_{Ds}=R^2/{\cal D}_s\sim 1$ s and $t_{Do}=R^2/{\cal D}_o\sim 500$ s, respectively. 

The time scales mentioned above can be observed when analyzing the maximum surfactant concentration $\Gamma_{\textin{max}}$ and mean surfactant concentration $\overline{\Gamma}$. The surfactant adsorbed onto the interface at $t=0$ s is swept toward the rear of the bubble very fast. At $t=0.008$ s, when the bubble has traveled a distance smaller than its radius (Fig.\ \ref{evolution}a), the maximum surfactant concentration $\Gamma_{\textin{max}}$ has increased over five times (Fig.\ \ref{evolution}d). However, the mean surfactant concentration $\overline{\Gamma}$ has remained almost constant (Fig.\ \ref{evolution}c). This confirms that surfactant convection dominates the process on the time scale $t_c=R/v_{zc}\sim 10^{-2}$ s. 

\begin{figure*}[hbt]
\begin{center}
\resizebox{0.3\textwidth}{!}{\includegraphics{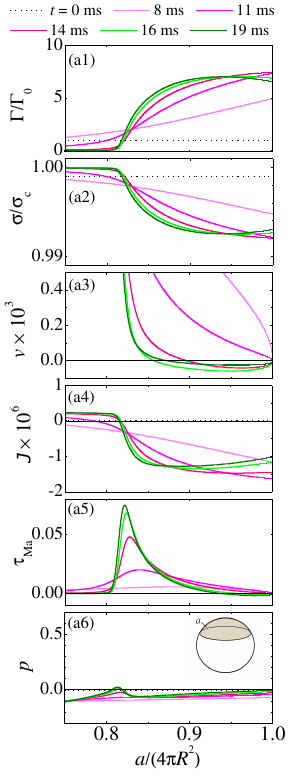}}\resizebox{0.3\textwidth}{!}{\includegraphics{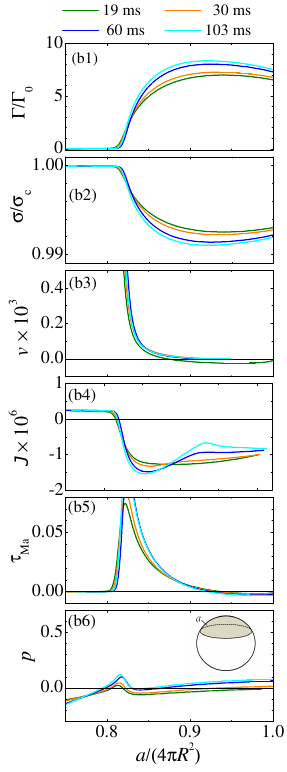}}\resizebox{0.29\textwidth}{!}{\includegraphics{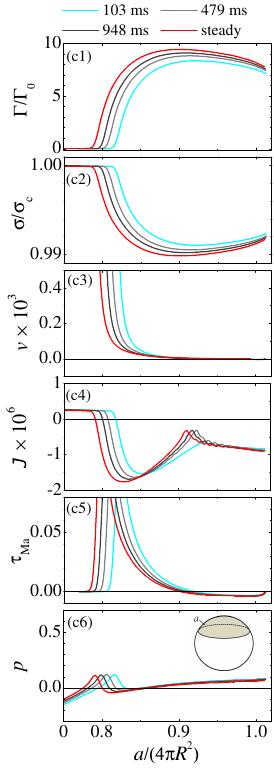}}
\end{center}
\caption{Surfactant surface concentration $\Gamma$, surface tension $\sigma$, surface velocity $v$ and net sorption flux ${\cal J}={\cal J}_a-{\cal J}_d$, Marangoni stress $\tau_{\textin{Ma}}$, and hydrostatic pressure $p$ as a function of $a$ for $R=0.66$ mm and $c_{\infty}=8\times 10^{-4}$ mol/m$^3$ of SDS. The flux and stresses are measured in terms of $v_{\sigma\mu}\Gamma_{\infty}/R$ and $\rho^{(o)} v_{\sigma\mu}^2$, respectively, where $v_{\sigma\mu}=\sigma_0/\mu^{(o)}$ is the visco-capillary velocity.} 
\label{Profiles_timeZoom}
\end{figure*}

Now, we analyze the evolution of the surfactant surface distribution during the convection-dominated stage of the process. For this purpose, we have re-plotted in Fig.\ \ref{Profiles_timeZoom} the profiles shown in Fig.\ \ref{Profiles_time}, separating them in several graphs and zooming into the rear of the bubble. The maximum surfactant concentration is initially located at the bubble bottom (see Fig.\ \ref{Profiles_timeZoom}-a1 for $t<14$ ms). The surface tension gradient produces a Marangoni stress $\tau_{\textin{Ma}}$ that significantly slows the surfactant-loaded part of the interface. This stress becomes large enough to reverse the interface motion in the bubble rear at $t\simeq 11$ ms (Fig.\ \ref{Profiles_timeZoom}-a3). The reverse flow transports the surfactant away from the bubble bottom, making $\Gamma_{\textin{max}}$ decrease even though there is net adsorption of surfactant, as indicated by the increase in $\overline{\Gamma}$ (Fig.\ \ref{evolution}c). At $t\simeq 19$ ms, the surfactant concentration profile exhibits the shape characteristic of the steady solution. Specifically, more than 80\% of the bubble surface is almost clean, and a thin diffusive boundary layer separates this region from the surfactant-loaded rear (Fig.\ \ref{Profiles_time}a). The Marangoni stress peak in the surface boundary layer practically immobilizes the surfactant-loaded part of the interface (Fig.\ \ref{Profiles_time}c).

The adsorption-desorption of surfactant controls the second phase of the dynamic adsorption layer growth. The flow drags the surfactant toward the rear of the bubble. The uneven surfactant concentration enhances adsorption in the almost-clean region and favors desorption in the loaded rear (Fig. \ref{Profiles_timeZoom}-b4). This results in a positive net sorption flux until the bubble reaches its steady velocity, as indicated by the continuous increase in $\overline{\Gamma}$ (Fig.\ \ref{evolution}c). 

The net sorption rate $\Phi$ [Eq.\ (\ref{Phi})] initially grows (Fig.\ \ref{evolution}e) as the clean part of the bubble surface enlarges. This occurs until the Marangoni stress peak almost immobilizes the surfactant-loaded region at $t\simeq 14$ ms (Fig.\ \ref{Profiles_timeZoom}-a3). Then, the bubble deformation increases the surface area (Fig.\ \ref{evolution}f). This allows for the growth of the surfactant-loaded region. This effect and the increase in the surfactant concentration favor desorption, reducing the net sorption rate. After the maximum velocity and deformation are reached ($t>0.1$ ms), the slight recovery of the spherical shape (Fig.\ \ref{Shape_evol}) contributes to enlarging the loaded area at the expense of the clean region (Fig.\ \ref{Profiles_timeZoom}-c1), which also reduces the net adsorption rate. The process continues until reaching the delicate balance between the surfactant convection, diffusion, and sorption corresponding to the steady regime \cite{RVCMLH24}.


\section{Conclusions}
\label{sec6}

We analyzed the transient bubble rising in the presence of a surfactant at very low concentrations. The experiments for $R=0.76$ mm and $c_{\infty}/c_{\textin{cmc}}\geq 5\times 10^{-4}$ of Triton X-100 showed that a helical instability eventually develops. The helical instability evolved toward a zig-zag motion as the Triton X-100 concentration increased. For the same relative concentrations $c_{\infty}/c_{\textin{cmc}}$ of Triton X-100 and SDS, Triton X-100 produces a smaller effect on a relatively short time scale, but this effect eventually exceeds that produced by SDS at sufficiently large times. The Triton X-100 effects are larger than those caused by SDS at the same absolute concentration. The magnitude of the surfactant effect increases as the bubble radius decreases for both Triton X-100 and SDS. The experiments allowed us to determine the parameter conditions that lead to an axisymmetric, straight (stable) bubble rising. The bubble motion remained stable for $R=0.66$ mm and the concentration $c_{\infty}=8\times 10^{-4}$ mol/m$^3$ of SDS. The experimental results satisfactorily agreed with the transient numerical solution for this case.

We numerically analyzed the growth of the dynamic adsorption layer of SDS for $R=0.66$ mm and $c_{\infty}=8\times 10^{-4}$ mol/m$^3$. The vertical velocity of a bubble in clean water increases monotonously. Conversely, the bubble velocity in our simulation attains a maximum at around $t\simeq 0.1$ s. This occurs despite the tiny surfactant concentration considered in our analysis, which suggests that the so-called ``overshooting"\ phenomenon occurs for any surfactant concentration. Both the Marangoni stress confined in the boundary layer and the increase in the low-pressure region area are responsible for the bubble deceleration. The bubble surface slightly expands during the bubble acceleration and compresses during the deceleration. A non-monotonous variation of the aspect ratio accompanies this expansion/compression. The amount of surfactant in the monolayer continuously increases over the bubble motion. 

One can distinguish four consecutive phases in the formation of the dynamic adsorption layer: the almost instantaneous convection of surfactant to the rear part of the bubble, the process influenced by the adsorption-desorption of surfactant, and a last and long process in which diffusion comes into place to balance convection and sorption. The evolution of the surfactant surface concentration is consistent with the time scales of the above-mentioned phases derived from a simple scaling analysis. We explained the non-monotonous behavior of the maximum surface concentration in terms of the reverse flow in the rear of the bubble right after the bubble release. We also determined the role played by the bubble's instantaneous deformation in the non-monotonous behavior of the net sorption rate.

\appendix

\section{Adsorption of surfactant to a clean bubble surface}

This appendix analyzes the time evolution of the surface coverage when an initially clean spherical bubble remains at rest in a liquid bath containing a surfactant. We consider the Henry model $\Gamma=k_a c/k_d$ in the limits of surfactant transport controlled by sorption kinetics and diffusion \citep{MS20}.

When surfactant transport is limited by sorption kinetics, diffusion in the bulk is assumed to be instantaneous, and the surfactant surface concentration $\Gamma_{\textin{kin}}$ obeys the exponential relaxation law \citep{MS20}
\begin{equation}
\frac{\Gamma_{\textin{kin}}}{\Gamma_{\textin{eq}}}=1-e^{-k_d t},
\end{equation}
where $\Gamma_{\textin{eq}}$ is the surface concentration at equilibrium. As can be observed, kinetically limited mass transfer is independent of bubble size, and $\tau_k=1/k_d$ is the characteristic time scale.

When adsorption is controlled by diffusion, equilibrium is assumed between the surface concentration and the bulk concentration at the surfactant sublayer. In this case, the surfactant surface concentration $\Gamma_{\textin{diff}}$ obeys the equation \citep{JBS04,AWA10,MS20}
\begin{equation}
\label{dif}
\frac{\Gamma_{\textin{diff}}}{\Gamma_{\textin{eq}}}=1+\ffrac{1}{\beta-\alpha}\left[\alpha e^{\alpha^2 t}\text{erfc}(\alpha\sqrt{t})-\beta e^{\beta^2 t}\text{erfc}(\beta\sqrt{t})\right],
\end{equation}
where 
\begin{equation}
\alpha=\frac{\sqrt{\cal D}_o}{2L_d}\left(1+\sqrt{1-4\Lambda_d}\right), \quad \beta=\frac{\sqrt{\cal D}_o}{2L_d}\left(1-\sqrt{1-4\Lambda_d}\right),
\end{equation}
$L_d=k_a/k_d$ is the so-called depletion length and $\Lambda_d=L_d/R$. As can be observed, diffusion-limited mass transfer depends on the bubble size. It can be seen that Eq.\ (\ref{dif}) is approximately 
\begin{equation}
\frac{\Gamma_{\textin{diff}}}{\Gamma_{\textin{eq}}}=1-e^{{\cal D}_o t/L_d^2}\, \text{erfc}(\sqrt{{\cal D}_o t}/L_d)
\end{equation}
for $\Lambda_d\ll 1$ and
\begin{equation}
\frac{\Gamma_{\textin{diff}}}{\Gamma_{\textin{eq}}}=1-e^{{\cal D}_o t/(L_d R)}
\end{equation}
for $\Lambda_d\gg 1$. Therefore, the corresponding time scales are $\tau_D=L_d^2/{\cal D}_o$ and $\tau_D=L_d R/{\cal D}_o$.

\begin{figure*}[hbt]
\begin{center}
\resizebox{0.38\textwidth}{!}{\includegraphics{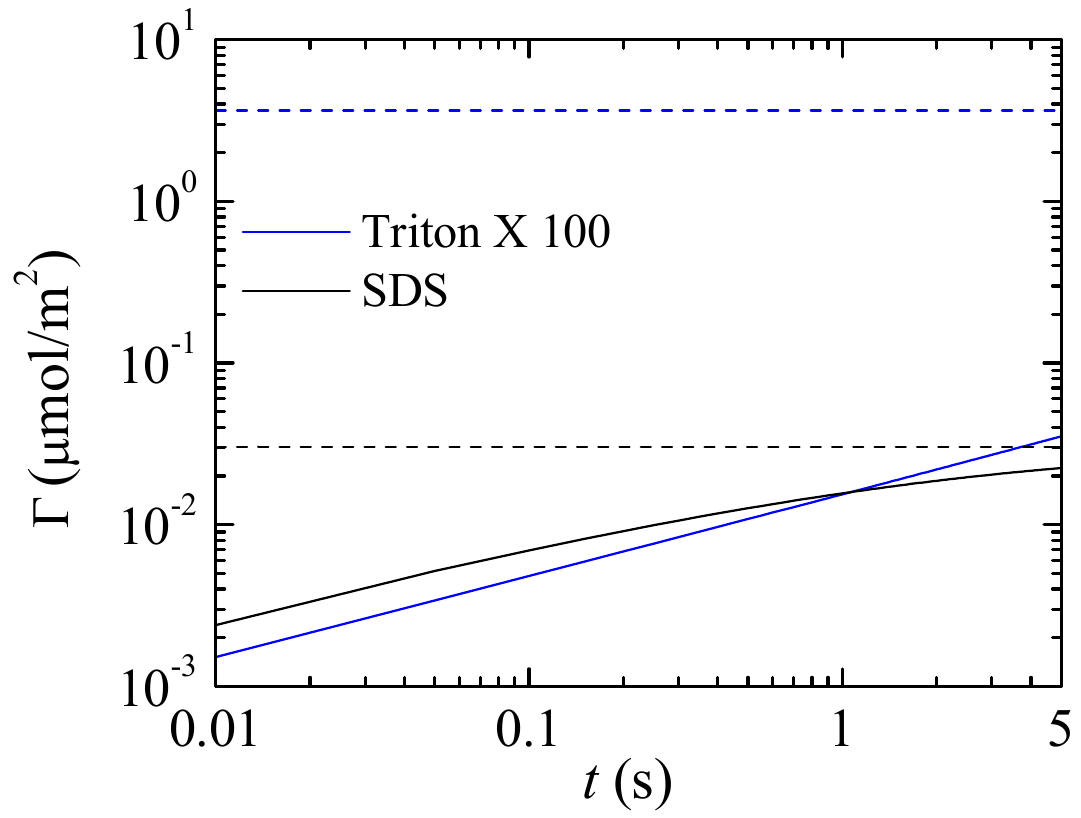}}\resizebox{0.38\textwidth}{!}{\includegraphics{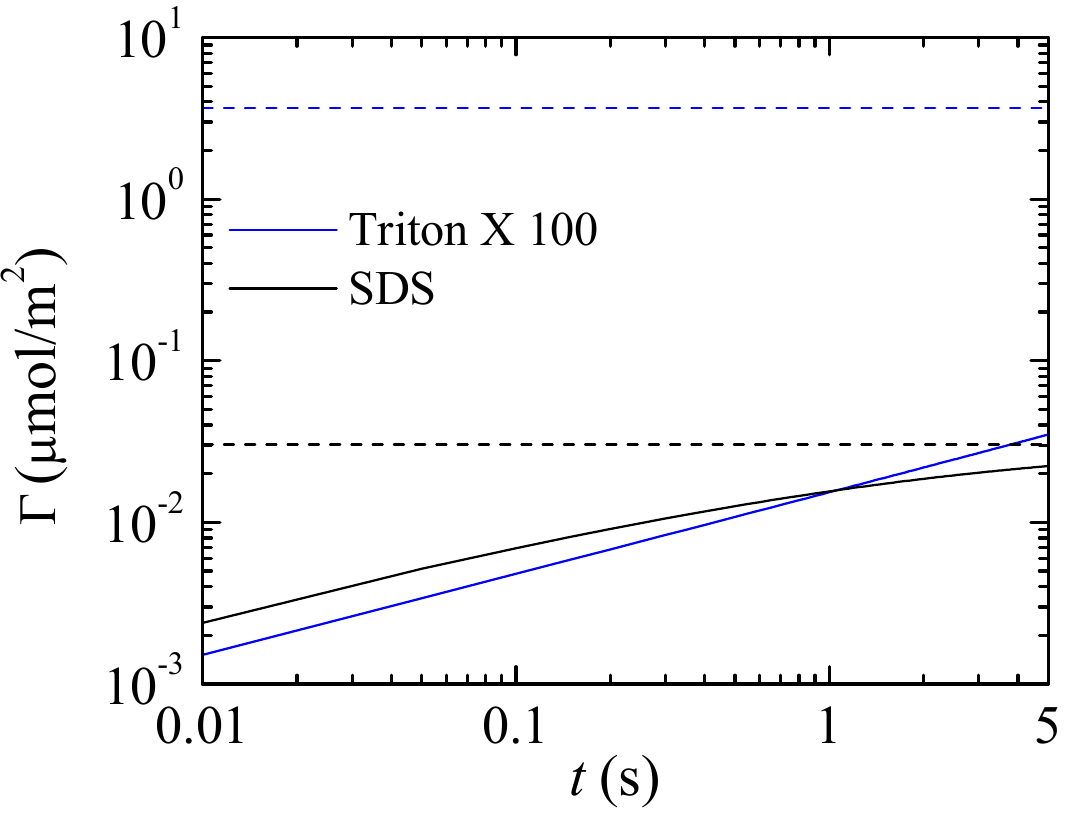}}
\end{center}
\caption{Surface surfactant concentration $\Gamma(t)$ for $c_{\infty}=8.3\times 10^{-4}$ mol/m$^3$ and $R=0.66$ mm (left) $R=0.76$ mm (right). The dashed lines are the values at equilibrium.} 
\label{difu}
\end{figure*}

The Damkohler number Da$=\tau_D/\tau_k$ takes values of the order of $10$ and $10^2$ for SDS and Triton X-100, respectively, indicating that the surfactant transfer to the bubble is limited by diffusion in both cases. Figure \ref{difu} shows the surfactant surface concentration $\Gamma(t)$ given by Eq.\ (\ref{dif}) for $c_{\infty}=8.3\times 10^{-4}$ mol/m$^3$. The influence of the bubble radius on $\Gamma(t)$ is negligible for both SDS and Triton X-100.

\vspace{1cm}

We gratefully acknowledge support from the Spanish Ministry of Science and Education (grant no. PID2022-140951OB-C21 and PID2022-140951OB-C22/AEI/10.13039/501100011033/FEDER) and Gobierno de Extremadura (grant no. GR21091). 


%

\end{document}